\documentclass[useAMS,usenatbib]{mn2e}
\usepackage[pdftex]{graphicx}

\def\araa{{ARA\&A}}%
\def\apj{{ApJ}}%
\def\apjl{{ApJ}}%
\def\ao{{Appl.~Opt.}}%
\def\aap{{A\&A}}%
\def\azh{{AZh}}%
\def\mnras{{MNRAS}}%
\def\nat{{Nature}}%
 \def\farcs@apj{%
  \mbox{%
   \kern  0.13ex.%
   \kern -0.95ex\raisebox{.9ex}{\scriptsize$\prime\prime$}%
   \kern -0.1ex%
  }%
 }%
\def\spose#1{\hbox to 0pt{#1\hss}}
\def\lta{\mathrel{\spose{\lower 3pt\hbox{$\mathchar"218$}}
     \raise 2.0pt\hbox{$\mathchar"13C$}}}
\def\gta{\mathrel{\spose{\lower 3pt\hbox{$\mathchar"218$}}
     \raise 2.0pt\hbox{$\mathchar"13E$}}}
\def\be{\begin{equation}}
\def\ee{\end{equation}}
\def\bea{\begin{eqnarray}}
\def\eea{\end{eqnarray}}

\def\bkappa{\pmb{\kappa}}
\def\bxi{\pmb{\xi}}
\def\md{{\rm d}}

\def\btheta{\pmb{\theta}}

\title[Anisotropic scattering models]{Power-law models of totally anisotropic scattering}
\author[A.V. Tuntsov, H.E. Bignall, M.A. Walker]{A.V. Tuntsov$^{1,2}$\thanks{E-mail: tyomichl@gmail.com},
H.E. Bignall$^3$\thanks{$\qquad\qquad\!\!\!$h.bignall@curtin.edu.au}, M.A. Walker$^1$\thanks{$\qquad\qquad\!\!\!$Mark.Walker@manlyastrophysics.org},\\
$^{1}$Manly Astrophysics, 3/22 Cliff St,  Manly 2095, Australia\\
$^{2}$Moscow M V Lomonosov State University, Sternberg Astronomical Institute, Moscow 119992, Russia\\
$^{3}$International Centre for Radio Astronomy Research, Curtin University, GPO Box U1987, Perth,
WA 6845, Australia}
\begin{document}

\date{Accepted 2012 November 30.  Received 2012 November 22; in original form 2012 September 21}

\pagerange{\pageref{firstpage}--\pageref{lastpage}} \pubyear{2012}

\maketitle

\label{firstpage}

\begin{abstract}
The interstellar scattering responsible for pulsar parabolic arcs, and for intra-day variability of compact radio quasars, is highly anisotropic in some cases. We numerically simulate these observed phenomena using totally anisotropic, power-law models for the electron density fluctuations which cause the scattering. By comparing our results to the scattered image of  PSR B0834+06 and, independently, to dual-frequency light curves of the quasar PKS1257-326, we constrain the nature of the scattering media on these lines of sight. We find that models with spectral indices slightly below $\beta=3$, including the one-dimensional Kolmogorov model, are broadly consistent with both data sets.  We confirm that a single physical model suffices for both sources, with the scattering medium simply being more distant in the case of B0834+06. This reinforces the idea that intra-day variability and parabolic arcs have a common cause in a type of interstellar structure which, though obscure, is commonplace. However, the implied gas pressure fluctuations are large compared to typical interstellar pressures, and the magnetic stresses are much larger still. Thus while these scattering media may be commonplace, their underlying dynamics appear quite extraordinary.
\end{abstract}

\begin{keywords}
Scattering -- turbulence -- ISM: general -- ISM: structure -- pulsars: individual: B0834+06 -- BL Lacertae objects: individual: PKS~1257-326
\end{keywords}

\section{Introduction}
\label{section:intro}

Radio waves are scattered by inhomogeneities in the ionised interstellar medium (ISM). The appearance of astronomical objects is thus altered and the ISM leaves its imprint in the received radiation. The effects of propagation include angular broadening and wander of images, variation in the arrival time of impulses, and modulations of the flux density of compact radio sources (i.e. scintillations). These various effects provide powerful tools with which to study the ISM, including exploration of small-scale features that cannot be seen with any other methods (see, for example,  the various contributions to \citealt{haverkorngoss2007}).

Although many aspects of the available data can be successfully described using a model based on distributed Kolmogorov turbulence in the ISM, there are also some significant departures from this picture \citep{rickett1990}. Perhaps the most striking examples are the Extreme Scattering Events \citep{fiedleretal1987, fiedleretal1994}, which appear to require discrete, high-density plasma concentrations of size $\sim10^{14}\,{\rm cm}$. And there are two other phenomena which are indicative of highly localised scattering regions dominating the rest of the line-of-sight, namely the large-amplitude intra-day variations (IDV) of some compact radio quasars \citep{KCJ1997, dennettthorpedebruyn2000, bignalletal2003}, and the parabolic arcs seen in the power-spectra of the dynamic spectra of radio pulsars \citep{stinebringetal2001}. At present it is unclear how  these three phenomena relate to the broader ISM, or how much they contribute to the scattering on a ``typical'' line-of-sight.

There is strong evidence for high levels of anisotropy in the scattering material responsible for IDV. These indications come from the quasi-sinusoidal character of the light-curves, and from modelling of the annual variation in scintillation time-scales, along with the scintillation pattern arrival time delays between widely separated telescopes  (Rickett, Kedziora-Chudczer \&  Jauncey 2002; \citealt{dennettthorpedebruyn2003, bignalletal2006}). The anisotropy, as measured by the major- to minor-axis ratio, $A$, must be $A\gg1$, and it is possible that $A\gg10$. In this circumstance totally anisotropic, one-dimensional scattering models are of interest as a simple, limiting case which can provide a good approximation to the actual circumstance. It was demonstrated by Walker, de~Bruyn and Bignall (2009) that a one-dimensional ($A\rightarrow\infty$) kinematic model offers a good explanation for the two-station time-delays and annual cycle of the scintillation time-scale of the very rapid IDV quasar J1819+3845.

In the case of pulsar parabolic arcs, highly anisotropic scattering models are known to give arcs which are very sharply defined \citep{cordesetal2006, walkeretal2004}. And a quasi-one-dimensional interpretation of the pulsar data is supported by observational studies which are directly sensitive to the angular structure of the scattered image \citep{walkeretal2008, briskenetal2010}.

In the present paper we explore the constraints that can be placed on the parameters of a one-dimensional, power law model of electron density fluctuations, in the context of IDV and parabolic arcs. We do so using extensive numerical simulations, which are compared against observations --- visually, and using summary statistics. We prefer this approach to any comparison based on ensemble-average models because the data are acquired over short intervals, which are far from the ensemble-average regime \citep{goodmannarayan1989}.

Section~\ref{section:model} introduces the model and simulation technique. Section~\ref{section:pulsar} derives model constraints from the observed image of a highly scattered pulsar, B0834+06,  and section~\ref{section:idv} formulates model constraints from the dual-frequency light curves of the IDV blazar PKS1257-326.  Section 5 summarises our constraints, and in section 6 we draw inferences concerning the nature of the scattering media.

\section{Scattering model and simulations}
\label{section:model}

We assume that there is a single thin screen \citep{bramley1954}, at an effective distance $D_\mathrm{eff}$, that dominates the scattering, as appropriate for both objects of interest \citep{hilletal2005, bignalletal2006}. We model the electron column density, $\kappa_e$, fluctuations as Gaussian random variables which are spatially correlated, with the power spectrum of $\kappa_e$ being \citep{leejokipii1976, narayangoodman1989}
\bea\label{Pkappaespectrum}
P(Q)=\frac{\sqrt\pi\,\Gamma\!\left(\beta/2\right)}{\lambda^2 r_e^2l_F^{\beta-1}}\frac{\alpha\, e^{-Q/Q_i}}{\left(Q^2+Q_0^2\right)^{\beta/2}}.
\eea
The spectrum of phase fluctuation $\phi=\lambda r_e\kappa_e$ follows trivially. In equation~(\ref{Pkappaespectrum}), $\lambda$ is the wavelength of observations, $r_e$  is the classical electron radius and $l_F=\sqrt{\lambda D_\mathrm{eff}/2\pi}$ is the Fresnel length. The parameters $\alpha$, $\beta$, $Q_0$ and $Q_i$ describe the spectrum itself. The quantity $\alpha$ is, up to a factor $\mathcal{O}(1)$, the square of the characteristic phase variation at a transverse separation equal to the Fresnel scale and thus it gauges the amplitude of scintillations (e.g. \citealt{narayan1992}): $\alpha\ll1$ indicates weak scintillations, whereas $\alpha\gg1$ corresponds to order-unity flux variations. Consequently we refer to $\alpha$ as the scattering strength. At sufficiently strong scattering, for power-law spectra with vanishing inner scale, $\alpha\approx u_R^{\beta-1}$ in terms of the $u_R$ parameter defined by equation~2.5 of \cite{rickett1990}.

Spectra with slopes $\beta<3$ will be referred to as ``shallow'' while those with $\beta>3$ as ``steep''. The classical theory of turbulent cascades \citep{kolmogorov1941} predicts a shallow slope of $\beta=8/3$, in the one-dimensional case (i.e. two-dimensional, but projected onto the plane of the sky), for the inertial range of a turbulent cascade. We emphasise, however, that the physics appropriate to interstellar scattering media is far from settled, either observationally \citep{blandfordnarayan1985, goodmannarayan1985, rickett1990, marongoldreich2001} or theoretically \citep{iroshnikov1963, kraichnan1965, goldreichsridhar1995,  boldyrev2006}. And this point is underlined by our current lack of understanding of the particular phenomena under consideration here. In this paper we consider a range of both shallow and steep spectra, $2<\beta<5$. 

The outer scale $Q_0^{-1}$ corresponds roughly to the largest scales at which the phase fluctuations remain correlated, and this scale dominates the statistics of phase, especially for steep slopes. Estimates for the value of $Q_0$ on different lines-of-sight vary considerably (\citealt{rickett1990}, Armstrong, Rickett \& Spangler 1995, \citealt{chepurnovlazarian2010}). Since the outer scale cannot exceed the size of individual strongly-scattering plasma patches, intermittency of the IDV phenomenon sets an upper limit on the outer scale of the media responsible for that phenomenon. Intermittency on a time-scale of months was seen in the intra-hour variable PKS0405-385 \citep{kedziora2006}. Similarly, after a decade of rapid variations J1819+3845 ceased to scintillate \citep{cimo2008}. Finally, amongst a sample of hundreds of compact radio sources when IDV was seen it was often intermittent on a timescale of months \citep{masiv2}. These observations indicate that  $Q_0^{-1}\lta 10^{14}\,\mathrm{cm}$. 

The inner scale, $Q_i^{-1}$, is that of the energy dissipation in the classical theory. The microscopic processes which may be responsible for such dissipation include ion-neutral collisions and Landau damping \citep{spangler1991}. The inner scale affects strongly the phase derivatives and small-scale correlation properties of the screen \citep{spanglergwinn1990}. A recent estimate, based on the impulse-response function for strongly-scattered pulsars \citep{rickettetal2009} puts the inner scale below $10^7\,{\rm cm}$ although similar considerations yielded values close to $10^8\,{\rm cm}$ \citep{bhatetal2004} earlier, and inner scales as large as $10^{11}\;{\rm cm}$ have previously been suggested \citep{colesetal1987}. To date there has been no determination of the inner scale specific to the highly anisotropic scattering media responsible for IDV or parabolic arcs. 

We simulate realisations of the electric field envelope, $u$, in the manner of \citet{martinflatte1988, narayangoodman1989}. In modelling PSR~B0834+06, a large spatial dynamic range $N\gta10^7$ is needed in order to keep edge effects from corrupting the simulated observations. For one-dimensional simulations, this is feasible, which allowed us to repeat the simulations on demand, build statistically meaningful samples, and explore the parameter space in an adaptive manner. When extracting light curves, we computed the squared magnitude of the field at the sampling point, $F_\nu=|u|^2$. For IDV sources, incoherent averaging was performed as well -- i.e. smoothing out the flux densities $F_\nu(R)$ -- with a Gaussian source profile. When statistical measures of the light curves were used for comparing data to the simulations -- e.g. auto- and cross-correlation functions (ACF and CCF) -- procedures were applied to the simulated flux densities which exactly paralleled the treatment of the comparison data. The angular structure of the images was estimated by Fourier transforming the spatial sampling of the field, so as to match the way the single dish data were analysed, then calculating its squared magnitude, $I(\theta)=|\int \md x\,u_\nu(x)\exp(2\pi i \nu x\theta/c)|^2$.

\begin{figure}
\center
\includegraphics[width=84mm]{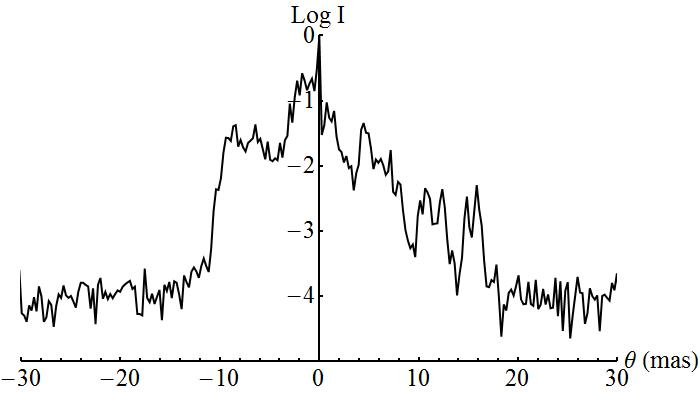}
\caption{The scattered image intensity $I$ as a function of angular distance $\theta$ of PSR B0834+06, determined from the delay-Doppler image of Walker et al (2008). The low-intensity baseline of this image, at $I\sim10^{-4}$, is just the instrumental noise-floor.}
\end{figure}

Altogether, we have five parameters in the model: strength $\alpha$, slope $\beta$, outer and inner scales $Q_0$ and $Q_i$ and the source intrinsic size $\sigma_\theta$ -- or, equivalently, its brightness temperature $T_b$, as per $F_\nu=2\pi\sigma_\theta^2 kT_b/\lambda^2$. Exploring the full five-dimensional space of parameters is not practicable and, as will be clear from the analysis below, is not warranted by the data. We neglect the intrinsic size of the pulsar emission region ($T_b\rightarrow\infty$), and for PKS1257-326 we consider only the case $T_b=10^{12}\,{\rm K}$ (in the observer frame) as representative of the brightness temperature of a compact synchrotron source. We are also able to fix the scattering strength, $\alpha$, for each set $(\beta, Q_i, Q_0)$ from the basic requirement on the image size or variability amplitude, which proved to be relatively stable from one realisation to another. Of the remaining three parameters, the inner  and outer scales have a restricted scope in that the theoretical appeal of a power-law is its scale-free nature. Our initial approach is therefore to set the inner scale to below our numerical resolution ($10^8\,\mathrm{cm}$ or $10^6\,\mathrm{cm}$ for PSR or IDV simulations, respectively), the outer scale to above the extent of the simulated domain ($10^{15}\;{\rm cm}$ or $10^{13}\;{\rm cm}$), and thus proceed to constrain $\beta$. Subsequently we allow some flexibility in either $Q_i$ or  $Q_0$, as appropriate, to see how our constraints relax.

\begin{figure}
\includegraphics[width=84mm]{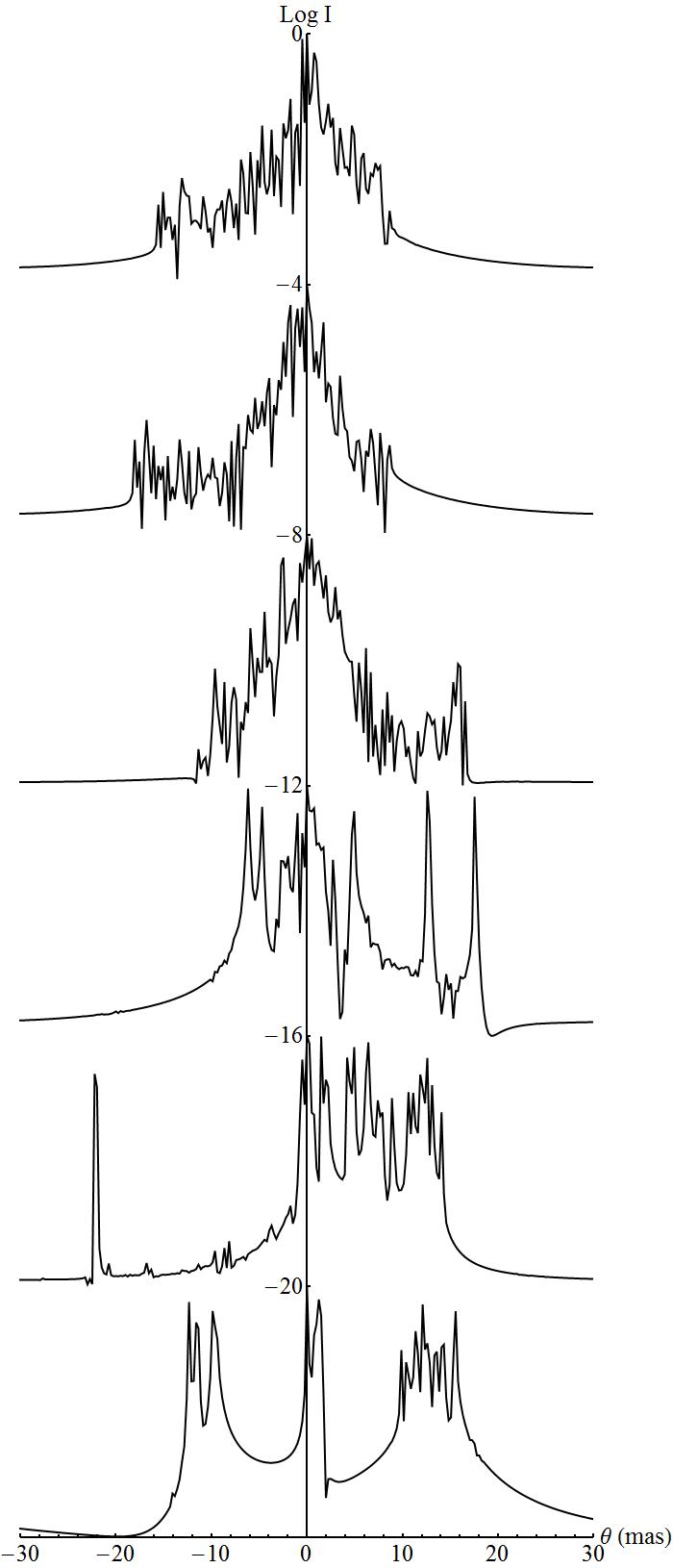}
\caption{Simulated scattered images of PSR B0834+06 for scale-free spectra with spectral indices (top to bottom): $\beta=2,2.5,3,3.5,4,4.5$. Each image has a constant intensity of $10^{-4}$ added, to mimic the instrumental noise-floor. Tthe images are offset from each other by four decades along the vertical axis.}
\label{fig:tyomafigure2}
\end{figure}

\section{PSR~B0834+06 constraints}\label{section:pulsar}
The pulsar B0834+06 offers one of the best examples of the parabolic arc phenomenon \citep{stinebringetal2001, hilletal2005}. The phenomenon itself is understood as arising from a quadratic relationship between the geometric delay and the scattering angle, for a preferred direction on the sky \citep{cordesetal2006}. That preferred direction may be determined by the  effective transverse velocity vector, but in cases such as B0834+06 where the arcs are very sharply defined one infers that the preferred direction arises from a strong anisotropy in the scattering \citep{walkeretal2004, cordesetal2006}. This inference has been  confirmed by construction of the electric field image in the delay-Doppler plane (Walker et al 2008), and by interferometric imaging using very long terrestrial baselines \citep{briskenetal2010}.

By virtue of acquiring data simultaneously from several sites, \cite{briskenetal2010} were able to break some of the degeneracies which are present in single-dish data and thus determine, for example, the distance of the scattering material along the line-of-sight ($D_{scr}=415\;{\rm pc}$); we adopt their measured screen distance.

\subsection{Data and simulations}

To constrain our model of the scattering medium we use the electric-field image of \cite{walkeretal2008}. For the present purposes, where we're modelling the scattering as one-dimensional, the appropriate mapping of the delay-Doppler image onto one angular coordinate is trivial: one simply tallies the squared magnitudes of the scattered waves at a given Doppler-shift. The resulting image is shown in figure 1. With different physical approximations one can arrive at a different angular structure from the same delay-Doppler image; thus \citet{gaorickettcoles2010} and \citet{rickett2011} have obtained a two-dimensional rendering under the assumption that dispersive delays are negligible. Even so, the major part of their resulting image is approximately one-dimensional. 

The data shown in figure 1 form a single example of a scattered image; any hypothesised spectrum for the scattering medium must plausibly be able to reproduce this image. It is not necessary to seek a direct match between specific features in the data and in the simulations. Rather we seek a match between the general characteristics of the observed and simulated images. Specific aspects of the simulated images which we have paid attention to are: shape of the maximum (plateau/peak/bell-like {\it etc.\/}); shape and slope of the profile ``wings''; and the roughness of the profile (continuous/smooth/spiky {\it etc.\/}). We used the width of the image or its main component to fix $\alpha$ for each set $(\beta, Q_i, Q_0)$.

\subsection{Constraints from image profiles}

We first compared the data to simulated images for pure power-law inhomogeneity spectra (where the inner scale is unresolved and the outer scale is larger than the size of our simulated medium). Figure~\ref{fig:tyomafigure2} shows a series of images representative of simulations for various values of the spectral index, $\beta$, with $Q_0^{-1}=10^{15}\,{\rm cm}$ and a negligible inner scale. The shallowest slopes, $\beta<2.5$ produce  image profiles that are too ``triangular'', with the intensity trending steeply downward as one moves away from the origin. By contrast the observed profile (figure 1) exhibits multiple scales with, for example, a relatively flat region of the profile terminating abruptly at $\theta\simeq-10\;$mas.

Steep slopes, $\beta>3$ are not compatible with the data either -- the images they generate are too spiky or altogether fragmented in comparison with a fairly solid core to the observed image. Slopes $2.5\lta\beta\lta3$ appear generally compatible with the data. Note that we are simulating images in the `snapshot' regime of the \cite{narayangoodman1989} classification, where the telescope integration time is shorter than the diffractive time scale; this is appropriate for the data which we are comparing to. Previous work has shown that the data do not resemble the ``ensemble-average'' image expected for a Kolmogorov spectrum \citep{briskenetal2010}.

When the scale-free assumption is relaxed, both shallower and steeper slopes can be made compatible with the data by restricting the range of the power law with a larger inner or a smaller outer scale, respectively. In particular, $\beta=2$ requires that $Q_i^{-1}\gta10^{9}-10^{10}\,\mathrm{cm}$. For steep spectra we find that $\beta=3.5$ calls for $Q_0^{-1}\lta10^{11}-10^{12}\,\mathrm{cm}$, whereas for $\beta=4$ an outer scale as low as $Q_0^{-1}\simeq 10^{10}\,\mathrm{cm}$ needs to be chosen for simulations to resemble the data.

We caution that the physical interpretation of our model (eq.~\ref{Pkappaespectrum}) becomes more obscure as the inner scale, $Q_i^{-1}$, increases and/or the outer scale, $Q_0^{-1}$, decreases. For a meaningful power-law model of the density fluctuations we certainly require $Q_i^{-1}\ll Q_0^{-1}$ -- corresponding to the inertial range in the Kolmogorov theory, for example -- and this is satisfied for all our models. But we note that for $\beta>3$ models to be consistent with figure 1 the outer scale must be comparable with, or even below the Fresnel-scale, which is $l_F\simeq10^{11}\,$cm. Thus, although $\beta>3$ fluctuation spectra can be made consistent with figure 1, consistency can only be achieved by fundamentally altering the character of the spectrum to the point where it is no-longer steep in respect of its refractive ($Q^{-1}\gta l_F$) properties. The merit of $\beta>3$ models is therefore questionable. 

\subsection{Additional constraints from dispersive delays}

\begin{figure}
\center
\includegraphics[width=84mm]{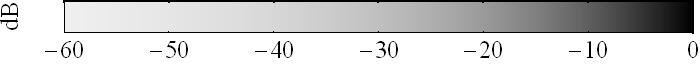}
\includegraphics[width=84mm]{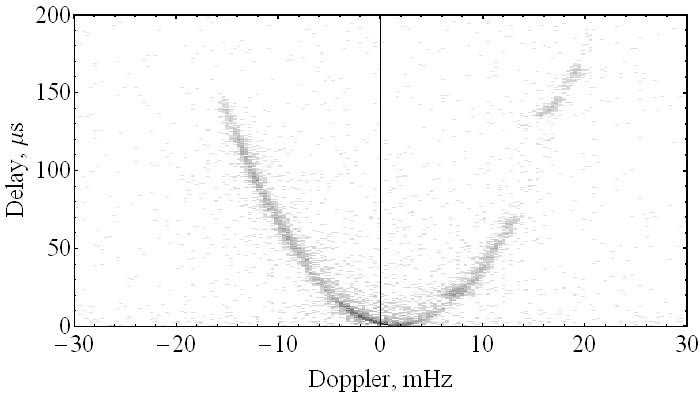}
\includegraphics[width=84mm]{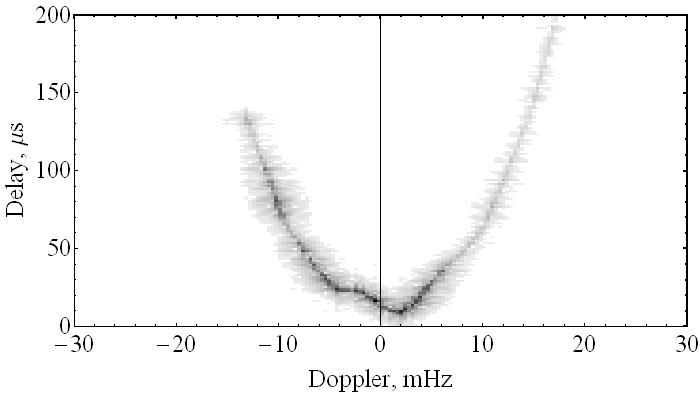}

\caption{Electric field magnitudes of the scattered radiation in delay vs. Doppler shift coordinates. The upper panel is the Walker~et~al.~(2008) construction from B0834+06 data, and the lower panel is a simulation with $\beta=3$, $Q_0^{-1}=10^{15}\,\mathrm{cm}$. For the B0834+06 data the locus of the power distribution conforms closely to a single parabolic arc, defined by the scattering geometry, whereas the simulation shows wiggles relative to the geometric arc. The wiggles are due to wave-speed (dispersive) delay contributions, and their absence in the data constrains the outer scale for steep fluctuation spectra. The intensity scale at the top is energy units relative to the maximum of a panel, $10\log |u(\tau, f_D)|^2/\max|u(\tau, f_D)|^2$.}

\label{fig:dds}

\end{figure}	

For models that produce acceptable image profiles, we also performed full dynamic-spectral simulations of the electric field. In some cases this revealed substantial deviations away from the notional parabolic locus of the power distribution in the delay-Doppler plane, as illustrated in the bottom panel of figure~\ref{fig:dds}. We attribute these ``wiggles'' to dispersive (wave-speed) delay contributions. The absence of any clear wiggles in the reconstruction of~\cite{walkeretal2008} (shown in the top panel of fig.~\ref{fig:dds}) allows us to constrain the outer scale to be $Q_0^{-1}\lta10^{13}\,\mathrm{cm}$ for $\beta=3$. For steeper fluctuation spectra the outer scale is already limited to very small values by matching to the observed image profile. For Kolmogorov, or shallower slopes, there is relatively little fluctuation power at low spatial frequencies and for these spectra our delay-Doppler analysis added no new information on the outer scale.

\smallskip\smallskip\noindent At this point it might be helpful to remind the reader that the specific inferences made in this section are restricted in scope to strictly one-dimensional models considered in this paper.

\section{IDV light-curve constraints}
\label{section:idv}

It is possible that the plasma inhomogenities responsible for pulsar parabolic arcs are of the same basic nature as those responsible for Intra-Day Variability, as both phenomena appear to involve highly concentrated regions of very strong, highly anisotropic scattering material. In \S5 we address that possibility. First, however, we consider what constraints can be placed on our one-dimensional power-law model~(eq. \ref{Pkappaespectrum}) from the IDV data alone.

\subsection{Data and simulations}

We utilised data for PKS1257-326 \citep{bignalletal2003}. This source is one of three bright, compact radio quasars which are known to exhibit large-amplitude radio variations on time-scales $\sim\;$hours. (The other two sources are PKS0405-385 [Kedziora-Chudczer et al 1997] and J1819+3845 [Dennett-Thorpe and de~Bruyn 2000].) The data we employ are simultaneous dual frequency ($4.8\,\mathrm{GHz}$ and $8.64\,\mathrm{GHz}$) light curves obtained with the Australia Telescope Compact Array (ATCA) as described in~\cite{bignalletal2003}, supplemented by similar data taken in 2003 as part of the same observational program.

We model the light curves assuming the Fresnel length of $l_F=3.3\times10^9\,\mathrm{cm}$ (in the X-band, as rescaled from the kinematical measurement of \citealt{walkerdebruynbignall2009}), implying the distance to the scatterer of $D\simeq 6.4\,\mathrm{pc}$. For simplicity, as the properties of the source are unknown in detail, we adopt equal angular sizes for the source in both C- and X-bands. (Assuming equal brightness temperatures instead does not affect the qualitative results derived below.) We use the observed amplitude of flux variation to fix $\alpha$ for each combination of  $(\beta, Q_i, Q_0)$. 

Following the approach adopted in the previous section, we did not try to match simulations to the data exactly, focusing instead on qualitative aspects of light curves. These are best summarised by the adjectives ``smooth'' and ``well-correlated'', which our simulated light-curves must reproduce. For source brightness temperatures $T_b\lta10^{13}\,\mathrm{K}$, as is expected for synchrotron radiation, the size of the source has a strong smoothing effect on the scintillation, and as a result the smoothness of the light-curves tells us little about the phase fluctuation spectrum (for $T_b$ in excess of $10^{14}\,{\rm K}$ the data strongly disfavour shallow spectra with inner scale below the Fresnel length). The degree and lag of inter-band correlation, on the contrary, prove useful to constrain the phase fluctuations at larger scales as the data exhibit very little prismatic shift.

\subsection{Constraints on prismatic shift}

\begin{figure}
\center
\includegraphics[width=84mm]{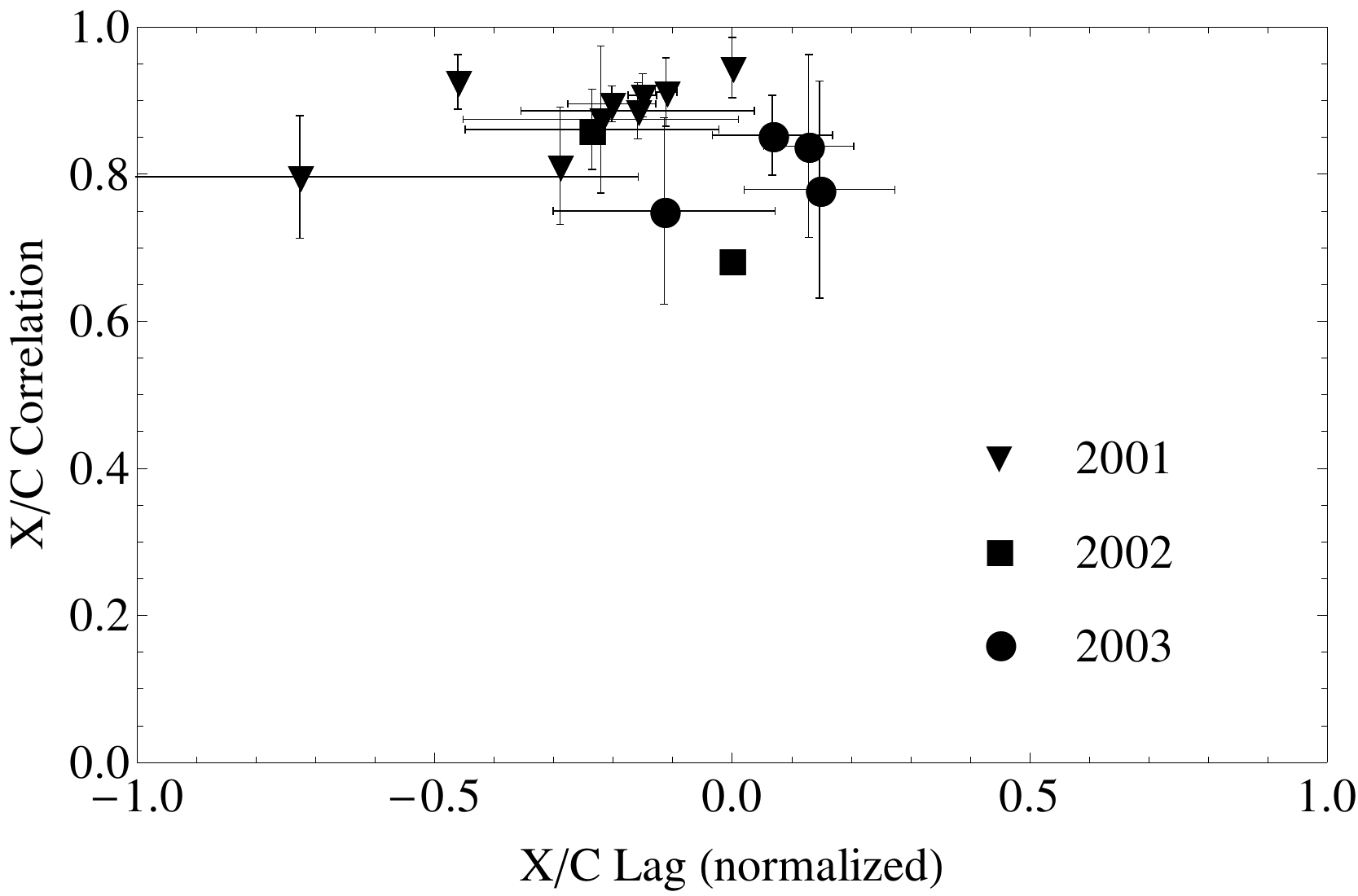}

\caption{Correlation coefficient of PKS~1257-326 C- and X-band light curves and normalised inter-band lag, $\hat\tau$ for three observational seasons. Positive lag means variations in C-band leading. For markers with error bars, the lag and correlation shown are averages for light curves from two or three consecutive dates.}

\label{fig:XCtaurnorm}

\end{figure}

The inter-band correlation statistics are summarised in Figure~\ref{fig:XCtaurnorm}, which shows the light curve pairs' correlation coefficient and inter-band lag normalised\footnote{Normalisation is introduced to take into account considerable annual variation in the effective velocity of the scattering screen.} by the overall variability timescale: $r\equiv\max_\tau \mathrm{CCF}_{X/C}(\tau)/(m_X m_C)$ and $\hat\tau\equiv\arg\max_\tau\mathrm{CCF}_{X/C}(\tau)/\sqrt{\mathrm{HWHM}_X\mathrm{HWHM}_C}$, where $m^2$ and $\mathrm{HWHM}$ are the height of the light curve $\mathrm{ACF}$ and half its width at the $m^2/2$ level. As one can see, the light curves are strongly correlated in the observed regime of moderately weak scattering, with the coefficient rarely dropping below $0.7$ and the lag being substantially below the temporal extent of a typical ``scintle'' -- i.e., the characteristic time scale of the observed quasi-sinusoidal variations of the flux density, quantified by the HWHM. 

The smallness of the observed lag strongly constrains the large-scale gradients, $\nabla\kappa_e$, of the electron column density, which are expected to induce prismatic shifts due to dispersion, i.e. dependence of the deflection angle, $\theta$, on the wavelength. At the observer, the difference in arrival angles translates into a lag of  $\tau_{X/C}\sim\Delta_{X/C}\theta\, D_\mathrm{eff}/v_\mathrm{eff}=(\lambda_C^2-\lambda_X^2)r_e\nabla\kappa_e D_\mathrm{eff}/2\pi v_\mathrm{eff}$. Since all gradient modes of scale greater than a few light-curve lengths, $\mathcal{T}$, contribute (incoherently, hence quadratures) to the prismatic shift, the typical lag, for a normal scale hierarchy $Q_i^{-1}\ll v_\mathrm{eff}\mathcal{T}\ll Q_0^{-1}$, is expected to be
\bea\label{tauexp}
\frac{\tau_{X/C}}{t_F\sqrt\alpha}\simeq \Theta_\beta \frac{\Delta\lambda^2}{\lambda_X^2}
\left[\begin{array}{ll}\left(t_F/\mathcal{T}\right)^{\frac{3-\beta}2}, &\beta<3 \cr \left(Q_0^{-1}/l_F\right)^{\frac{\beta-3}2}, &\beta>3\end{array}\right., 
\eea
with the coefficient $\Theta_\beta\sim\mathcal{O}(1)$ and $t_F\equiv l_F/v_\mathrm{eff}$. 

For PKS1257-326 at C- and X-band, the scattering is moderately weak, implying $\alpha^{1/2}\sim m\sim0.1$ and $t_F\sim$ duration of a scintle. Therefore for the prismatic shift to be small in Fresnel units we conclude from equation 2 that either (i) the slope of the fluctuation spectrum is shallow, or (ii) the slope is steep, but the outer scale is very small, $Q_0^{-1}\lta l_F\sim\mathrm{a~few}\times10^9\,\mathrm{cm}$. These conclusions are supported by our numerical simulations and are valid for various values of the source size around our assumed $T_b\sim10^{12}\;$K.

We note that the observed lags have previously been interpreted in terms of  an offset in the apparent centroids of the source in the C- and X-bands (\citealt{bignallhodgson2012, bignalletal2003}; Macquart et al. 2012).In this case, the measured values of $\tau_{X/C}$ should themselves be considered as upper limits on the prismatic shifts introduced by the scattering screen, and the constraints we have given are therefore conservative. 

\section{Summary of constraints}
We have constrained the form of the fluctuation spectrum (eq.~\ref{Pkappaespectrum}) independently for PSR B0834+06 and IDV PKS1257-326. However, there is some commonality between the two sets of constraints: neither spectrum is permitted to have much refractive fluctuation --- i.e. not much power in the region $l_FQ<1$. In the scale-free case ($Q_il_F\rightarrow\infty$, $Q_0l_F\rightarrow0$) that translates to $\beta<3$ for both scattering media. This constraint on $\beta$ relaxes if we admit a finite outer scale, but for both pulsar and IDV source the outer scale is forced to lie close to the Fresnel-scale in order to expand the permitted range of $\beta$ significantly on the steep side. We have already noted the questionable merit of such a description for PSR B0834+06, and with the same circumstance required for IDV PKS1257-326 we reject the model. The reason is simply that the Fresnel-scale for our IDV data is two orders of magnitude smaller than for our pulsar data: steep spectra can only be made to fit the data by contrivance.

Implicit in the above logic is the expectation that we can explain both PKS1257-326 and PSR B0834+06 with a single fluctuation spectrum; we now concentrate on that issue. The known points of similarity are as follows. First, both require very strongly scattering media. Second, both require high levels of anisotropy in the scattering media --- presumably enforced by a strong and ordered magnetic field. Third, there appears to be a high degree of spatial localisation for both phenomena, in that there are no known cases in which a given scatterer has been seen to affect more than a single background radio source. Fourth, as we have shown in sections 3 and 4, under the assumption of power-law fluctuations, both spectra are shallow, with $\beta\lta3$. And finally, the recent measurement (Brisken et al 2010) of the parallel (to the scattering angle) velocity component of the scattering medium towards PSR B0834+06 is similar to the results previously obtained for PKS1257-326 \citep{bignalletal2003} and J1819+3845 \citep{dennettthorpedebruyn2003}.

Our models allow us to reinforce the first of these points. The scattering strength inferred for PSR~B0834+06, when rescaled to the X-band and the distance of the PKS1257-326 scattering medium,  becomes $\alpha_{0834} (\lambda_{1257}/\lambda_{0834})^2 (l_{F, 1257}/l_{F,0834})^{\beta-1}\simeq 5.6\times10^4 (321/8640)^2(3.3\times10^9/2.3\times10^{11})^{\beta-1}\simeq 0.016$ (numerical value for the scale-free $\beta=3$ case). This is comparable to the actual $\alpha_{1257}\simeq 0.03$ we required to model the observed scintillations of PKS1257-326. The two values can be considered consistent with each other within the accuracy that the present analysis and data allow. In other words: both scattering phenomena can be attributed to the same type of physical structure, just placed at different distances from the observer. 

Given these points of similarity between the two scattering media, we consider it established that the scatterers are fundamentally similar.
Thus our preferred spectral index range for power-law inhomogeneities is $2.5\lta\beta\lta3$; this includes the one-dimensional Kolmogorov spectrum ($\beta=8/3$). At the uppermost end of this range the observed lack of wiggles in the B0834+06 parabolic arc constrains the outer scale to be $Q_0^{-1}\lta10^{13}\,\mathrm{cm}$. But for Kolmogorov, or shallower spectra these wiggles are too small to notice even when the outer scale is as large as permitted by intermittency of the IDV phenomenon (i.e. $Q_0^{-1}\lta10^{14}\,\mathrm{cm}$).

Slopes shallower than $\beta=2.5$ can be allowed by postulating a large inner scale, and our modelling does not rule out such possibility. However, similar comments apply in this case as to the interpretation of a steep spectrum with a very small outer scale. If a large inner scale is needed then we should consider the possibility that our model is inappropriate, even though it can yield results similar to the observations. For example, it has been proposed by \cite{briskenetal2010} that for B0834+06 the statistics of the electron-density fluctuations are non-uniform. If so then a great deal of freedom is permitted in the modelling, and even if the fluctuations are ``power-law'' none of that freedom is present in the model we have utilised here.

\section{Nature of the scattering media}
\label{section:discussion}

Unlike the strength  parameter, $\alpha$, the variance of the electron column density $\sigma_{\kappa_e}$ is not fixed by the data, because the integral of~(\ref{Pkappaespectrum}) is dominated by its lower limit and the outer scale is poorly constrained for our preferred, shallow slopes. Luckily, this is not the case for the variance of the electron volume density. Assuming the overall screen thickness $Z$ is of the same order as the outer scale, the electron volume density variance, for a generic orientation of the screen with respect to the line of sight, is 
\bea
\sigma^2_{n_e}\sim\frac{Q_0}Z\sigma^2_{\kappa_e}\simeq \frac{\pi\left[1+\mathcal{O}\left(Q_0/Q_i\right)\right]}{2\Gamma\left(\frac{3-\beta}2\right)\sin\frac\pi2(\beta-1)}\frac{\alpha Q_0^{3-\beta}}{\lambda^2r_e^2l_F^{\beta-1}},
\eea
which is very weakly dependent on $Q_0$ for $\beta$ close to 3. In particular, for the scale-free $\beta=3$ model the implied electron density r.m.s. is $\Delta n_e\sim30\,\mathrm{cm}^{-3}$, independent of the outer scale. And for the one-dimensional Kolmogorov spectrum this reduces to $\Delta n_e\sim10\,\mathrm{cm}^{-3}$ for the largest plausible outer scale ($10^{14}\,\mathrm{cm}$). This value is consistent with Rickett's (2011) estimate for B0834+06, based on the same data we have used.

Bearing in mind that we are dealing with ionised gas (temperature $\sim10^4\;$K), the foregoing estimates imply that the scattering medium manifests fluctuations in gas pressure which are an order of magnitude greater than typical pressures in the diffuse ISM: $P_{gas}\gg P_{ISM}$. This is puzzling, as one expects pressure equilibration on a sound-crossing time. This is a familiar problem in the context of Extreme Scattering Events (Romani, Blandford \& Cordes 1987), where the implied gas pressures are even larger than our estimate following equation~(3). But for the scattering media we have modelled, assuming that the anisotropy is due to a magnetic field, the gas pressure problem is only the tip of the iceberg.

The high anisotropy in the electron-density fluctuations indicates that the gas pressure fluctuations are small in comparison with the magnetic stress in the medium: $P_B\gg P_{gas}$. And the magnetic field must have its origins in a current system which is held in place by material stresses. The specific nature of the material -- call it ``X'' -- is unknown; but in order for the material stresses to at least balance the magnetic stresses we must have $P_X\gta P_B$. Thus we have
\bea
P_X\gta P_B \gg P_{gas} \gg P_{ISM}.
\eea
This ordering indicates that the diffuse ISM has little influence upon the scattering media we are modelling, and is practically irrelevant to the dynamics of the material, $X$, which is ultimately responsible for the observed phenomena. 

The conclusions we have just drawn rest on the assumption that $Z\sim Q_0^{-1}$. If, instead, $Z$ is much larger than the outer scale then smaller pressure fluctuations correspond to a given column-density variance. However, it is only possible to mitigate the stress hierarchy of equation 4, so that $P_X\sim P_{ISM}$, if $Z\gta10^6\,Q_0^{-1}$ --- i.e. an enormous geometric factor must be invoked. A further, more serious problem faces us if we try to overthrow the stress hierarchy of equation 4. If $P_B\sim P_{ISM}$, so that the magnetic stresses are no higher than typical interstellar stresses, then we do not expect the magnetic field to have a constant orientation through the scattering medium. But the fact that the observed anisotropy is $A\gg1$ tells us that the orientation is constant to within a projected angle of $\lta1/A\;$radian. For these reasons we consider ``ambient pressure'' models of the IDV and parabolic arc phenomena (e.g. \citealt{penking2012}) to be untenable.

\section{Summary and Conclusions}
We have numerically simulated radio-wave scattering by a totally anisotropic, power-law spectrum of electron-density fluctuations. Comparing our results to data on PSR B0834+06 and IDV PKS1257-326 we constrain the spectral index of the power-law to lie in the range $2.5\lta\beta\lta3$ --- a range which includes the one-dimensional Kolmogorov spectrum. The permitted range of $\beta$ can be expanded by introducing inner  and/or outer scales, but significant expansion requires such a large inner scale, and/or small outer scale, that  one or both scales are playing a major role, thus destroying the essence of the power-law. 

We find that the amplitude of the density fluctuations is comparable in the two distinct scattering regions which we model. This, together with the small spatial extent of the scattering media and their high levels of anisotropy, encourages the view that some pulsar parabolic arcs and the IDV of flat-spectrum quasars are due to the same types of scattering concentrations. The nature of these scatterers is obscure, but a simple-minded consideration of the associated physical stresses shows that there is a strong hierarchy, with the diffuse ISM at the lower end. We conclude that the diffuse ISM is largely irrelevant to these remarkable scintillation phenomena.

\section*{Acknowledgments}
MAW has benefited greatly from discussion of the physics with Barney Rickett and Bill Coles. We thank the referee for detailed comments and helpful suggestions.

\label{lastpage}


\begin{thebibliography}{}

\bibitem[\protect\citeauthoryear{{Bhat}, {Cordes}, {Camilo}, {Nice} \&
  {Lorimer}}{{Bhat} et~al.}{2004}]{bhatetal2004}
{Bhat} N.~D.~R.,  {Cordes} J.~M.,  {Camilo} F.,  {Nice} D.~J.,
  {Lorimer} D.~R.,  2004, \apj, 605, 759

\bibitem[\protect\citeauthoryear{{Bignall} \& {Hodgson}}{{Bignall} \&
  {Hodgson}}{2012}]{bignallhodgson2012}
{Bignall} H.~E.,  {Hodgson} J.~A.,  2012, in {Griffin} R.~E.~M.,  {Hanisch}
  R.~J.,   {Seaman} R.,  eds, IAU Symposium Vol.~285 of IAU Symposium, {On
  Rapid Interstellar Scintillation of Quasars: PKS 1257-326 Revisited}.
pp 129--132

\bibitem[\protect\citeauthoryear{{Bignall}, {Jauncey}, {Lovell}, {Tzioumis},
  {Kedziora-Chudczer}, {Macquart}, {Tingay}, {Rayner} \& {Clay}}{{Bignall}
  et~al.}{2003}]{bignalletal2003}
{Bignall} H.~E.,  {Jauncey} D.~L.,  {Lovell} J.~E.~J.,  {Tzioumis} A.~K.,
  {Kedziora-Chudczer} L.,  {Macquart} J.-P.,  {Tingay} S.~J.,  {Rayner} D.~P.,
    {Clay} R.~W.,  2003, \apj, 585, 653

\bibitem[\protect\citeauthoryear{{Bignall}, {Macquart}, {Jauncey}, {Lovell},
  {Tzioumis} \& {Kedziora-Chudczer}}{{Bignall} et~al.}{2006}]{bignalletal2006}
{Bignall} H.~E.,  {Macquart} J.-P.,  {Jauncey} D.~L.,  {Lovell} J.~E.~J.,
  {Tzioumis} A.~K.,    {Kedziora-Chudczer} L.,  2006, \apj, 652, 1050

\bibitem[\protect\citeauthoryear{{Blandford} \& {Narayan}}{{Blandford} \&
  {Narayan}}{1985}]{blandfordnarayan1985}
{Blandford} R.,  {Narayan} R.,  1985, \mnras, 213, 591

\bibitem[\protect\citeauthoryear{{Boldyrev}}{{Boldyrev}}{2006}]{boldyrev2006}
{Boldyrev} S.,  2006, Physical Review Letters, 96, 115002

\bibitem[\protect\citeauthoryear{{Bramley}}{{Bramley}}{1954}]{bramley1954}
{Bramley} E.~N.,  1954, Royal Society of London Proceedings Series A, 225, 515

\bibitem[\protect\citeauthoryear{{Brisken}, {Macquart}, {Gao}, {Rickett},
  {Coles}, {Deller}, {Tingay} \& {West}}{{Brisken}
  et~al.}{2010}]{briskenetal2010}
{Brisken} W.~F.,  {Macquart} J.-P.,  {Gao} J.~J.,  {Rickett} B.~J.,  {Coles}
  W.~A.,  {Deller} A.~T.,  {Tingay} S.~J.,    {West} C.~J.,  2010, \apj, 708,
  232

\bibitem[\protect\citeauthoryear{{Chepurnov} \& {Lazarian}}{{Chepurnov} \&
  {Lazarian}}{2010}]{chepurnovlazarian2010}
{Chepurnov} A.,  {Lazarian} A.,  2010, \apj, 710, 853

\bibitem[\protect\citeauthoryear{{Cim{\`o}}}{{Cim{\`o}}}{2008}]{cimo2008}
{Cim{\`o}} G.,  2008, in The role of VLBI in the Golden Age for Radio Astronomy
  {Two-Dimensional Time Delay Measurement of a fast Scintillator using VLBI
  arrays.}

\bibitem[\protect\citeauthoryear{{Coles}, {Rickett}, {Codona} \&
  {Frehlich}}{{Coles} et~al.}{1987}]{colesetal1987}
{Coles} W.~A.,  {Rickett} B.~J.,  {Codona} J.~L.,    {Frehlich} R.~G.,  1987,
  \apj, 315, 666

\bibitem[\protect\citeauthoryear{{Cordes}, {Rickett}, {Stinebring} \&
  {Coles}}{{Cordes} et~al.}{2006}]{cordesetal2006}
{Cordes} J.~M.,  {Rickett} B.~J.,  {Stinebring} D.~R.,    {Coles} W.~A.,  2006,
  \apj, 637, 346

\bibitem[\protect\citeauthoryear{{Dennett-Thorpe} \& {de
  Bruyn}}{{Dennett-Thorpe} \& {de Bruyn}}{2000}]{dennettthorpedebruyn2000}
{Dennett-Thorpe} J.,  {de Bruyn} A.~G.,  2000, \apjl, 529, L65

\bibitem[\protect\citeauthoryear{{Dennett-Thorpe} \& {de
  Bruyn}}{{Dennett-Thorpe} \& {de Bruyn}}{2003}]{dennettthorpedebruyn2003}
{Dennett-Thorpe} J.,  {de Bruyn} A.~G.,  2003, \aap, 404, 113

\bibitem[\protect\citeauthoryear{{Fiedler}, {Dennison}, {Johnston}, {Waltman}
  \& {Simon}}{{Fiedler} et~al.}{1994}]{fiedleretal1994}
{Fiedler} R.,  {Dennison} B.,  {Johnston} K.~J.,  {Waltman} E.~B.,    {Simon}
  R.~S.,  1994, \apj, 430, 581

\bibitem[\protect\citeauthoryear{{Fiedler}, {Dennison}, {Johnston} \&
  {Hewish}}{{Fiedler} et~al.}{1987}]{fiedleretal1987}
{Fiedler} R.~L.,  {Dennison} B.,  {Johnston} K.~J.,    {Hewish} A.,  1987,
  \nat, 326, 675

\bibitem[\protect\citeauthoryear{{Gao}, {Rickett} \& {Coles}}{{Gao}
  et~al.}{2010}]{gaorickettcoles2010}
{Gao} J.~J.,  {Rickett} B.~J.,    {Coles} W.~A.,  2010, in Society of
  Photo-Optical Instrumentation Engineers (SPIE) Conference Series Vol.~7800 of
  Society of Photo-Optical Instrumentation Engineers (SPIE) Conference Series,
  {Scattered image reconstruction of Pulsar B0834+06}

\bibitem[\protect\citeauthoryear{{Goldreich} \& {Sridhar}}{{Goldreich} \&
  {Sridhar}}{1995}]{goldreichsridhar1995}
{Goldreich} P.,  {Sridhar} S.,  1995, \apj, 438, 763

\bibitem[\protect\citeauthoryear{{Goodman} \& {Narayan}}{{Goodman} \&
  {Narayan}}{1985}]{goodmannarayan1985}
{Goodman} J.,  {Narayan} R.,  1985, \mnras, 214, 519

\bibitem[\protect\citeauthoryear{{Goodman} \& {Narayan}}{{Goodman} \&
  {Narayan}}{1989}]{goodmannarayan1989}
{Goodman} J.,  {Narayan} R.,  1989, \mnras, 238, 995

\bibitem[\protect\citeauthoryear{{Haverkorn} \& {Goss}}{{Haverkorn} \&
  {Goss}}{2007}]{haverkorngoss2007}
{Haverkorn} M.,  {Goss} W.~M.,  eds, 2007, {SINS - Small Ionized and Neutral
  Structures in the Diffuse Interstellar Medium} Vol.~365 of Astronomical
  Society of the Pacific Conference Series

\bibitem[\protect\citeauthoryear{{Hill}, {Stinebring}, {Asplund}, {Berwick},
  {Everett} \& {Hinkel}}{{Hill} et~al.}{2005}]{hilletal2005}
{Hill} A.~S.,  {Stinebring} D.~R.,  {Asplund} C.~T.,  {Berwick} D.~E.,
  {Everett} W.~B.,    {Hinkel} N.~R.,  2005, \apjl, 619, L171

\bibitem[\protect\citeauthoryear{{Iroshnikov}}{{Iroshnikov}}{1963}]{iroshnikov1963}
{Iroshnikov} P.~S.,  1963, \azh, 40, 742

\bibitem[\protect\citeauthoryear{{Kedziora-Chudczer}}{{Kedziora-Chudczer}}{2006}]{kedziora2006}
{Kedziora-Chudczer} L.,  2006, \mnras, 369, 449

\bibitem[\protect\citeauthoryear{{Kedziora-Chudczer}, {Jauncey}, {Wieringa},
  {Walker}, {Nicolson}, {Reynolds} \& {Tzioumis}}{{Kedziora-Chudczer}
  et~al.}{1997}]{KCJ1997}
{Kedziora-Chudczer} L.,  {Jauncey} D.~L.,  {Wieringa} M.~H.,  {Walker} M.~A.,
  {Nicolson} G.~D.,  {Reynolds} J.~E.,    {Tzioumis} A.~K.,  1997, \apjl, 490,
  L9+

\bibitem[\protect\citeauthoryear{{Kolmogorov}}{{Kolmogorov}}{1941}]{kolmogorov1941}
{Kolmogorov} A.,  1941, Akademiia Nauk SSSR Doklady, 30, 301

\bibitem[\protect\citeauthoryear{{Kraichnan}}{{Kraichnan}}{1965}]{kraichnan1965}
{Kraichnan} R.~H.,  1965, Physics of Fluids, 8, 1385

\bibitem[\protect\citeauthoryear{{Lee} \& {Jokipii}}{{Lee} \&
  {Jokipii}}{1976}]{leejokipii1976}
{Lee} L.~C.,  {Jokipii} J.~R.,  1976, \apj, 206, 735

\bibitem[\protect\citeauthoryear{{Lovell}, {Rickett}, {Macquart}, {Jauncey},
  {Bignall}, {Kedziora-Chudczer}, {Ojha}, {Pursimo}, {Dutka}, {Senkbeil} \&
  {Shabala}}{{Lovell} et~al.}{2008}]{masiv2}
{Lovell} J.~E.~J.,  {Rickett} B.~J.,  {Macquart} J.-P.,  {Jauncey} D.~L.,
  {Bignall} H.~E.,  {Kedziora-Chudczer} L.,  {Ojha} R.,  {Pursimo} T.,  {Dutka}
  M.,  {Senkbeil} C.,    {Shabala} S.,  2008, \apj, 689, 108

\bibitem[\protect\citeauthoryear{{Macquart}, {Godfrey}, {Bignall} \& {Hodgson}}{{Macquart} et~al.}{2012}]{macquartetal2012}
{Macquart} J.-P.,  {Godfrey} L.~E.~H., {Bignall} H.~E.,   {Hodgson} J.~A. , 2012 \apj, in press

\bibitem[\protect\citeauthoryear{{Maron} \& {Goldreich}}{{Maron} \&
  {Goldreich}}{2001}]{marongoldreich2001}
{Maron} J.,  {Goldreich} P.,  2001, \apj, 554, 1175

\bibitem[\protect\citeauthoryear{{Martin} \& {Flatte}}{{Martin} \&
  {Flatte}}{1988}]{martinflatte1988}
{Martin} J.~M.,  {Flatte} S.~M.,  1988, \ao, 27, 2111

\bibitem[\protect\citeauthoryear{{Narayan}}{{Narayan}}{1992}]{narayan1992}
{Narayan} R.,  1992, Royal Society of London Philosophical Transactions Series
  A, 341, 151

\bibitem[\protect\citeauthoryear{{Narayan} \& {Goodman}}{{Narayan} \&
  {Goodman}}{1989}]{narayangoodman1989}
{Narayan} R.,  {Goodman} J.,  1989, \mnras, 238, 963

\bibitem[\protect\citeauthoryear{{Pen} \& {King}}{{Pen} \&
  {King}}{2012}]{penking2012}
{Pen} U.-L.,  {King} L.,  2012, \mnras, 421, L132

\bibitem[\protect\citeauthoryear{{Rickett}}{{Rickett}}{2011}]{rickett2011}
{Rickett} B.,  2011, in {V.~Florinski, J.~Heerikhuisen, G.~P.~Zank, \&
  D.~L.~Gallagher } ed., American Institute of Physics Conference Series
  Vol.~1366 of American Institute of Physics Conference Series, {Anisotropic
  and Intermittent Turbulence in the Warm Ionized Interstellar Medium}.
pp 107--114

\bibitem[\protect\citeauthoryear{{Rickett}, {Johnston}, {Tomlinson} \&
  {Reynolds}}{{Rickett} et~al.}{2009}]{rickettetal2009}
{Rickett} B.,  {Johnston} S.,  {Tomlinson} T.,    {Reynolds} J.,  2009, \mnras,
  395, 1391

\bibitem[\protect\citeauthoryear{{Rickett}}{{Rickett}}{1990}]{rickett1990}
{Rickett} B.~J.,  1990, \araa, 28, 561

\bibitem[\protect\citeauthoryear{{Rickett}, {Kedziora-Chudczer} \&
  {Jauncey}}{{Rickett} et~al.}{2002}]{rickett2002}
{Rickett} B.~J.,  {Kedziora-Chudczer} L.,    {Jauncey} D.~L.,  2002, \apj, 581,
  103

\bibitem[\protect\citeauthoryear{{Romani}, {Blandford} \&
  {Cordes}}{{Romani} et~al.}{2002}]{romanietal1987}
{Romani} R.~W.,  {Blandford} R.~D.,  {Cordes} J.~M.,  1987, \nat, 328, 324

\bibitem[\protect\citeauthoryear{{Spangler}}{{Spangler}}{1991}]{spangler1991}
{Spangler} S.~R.,  1991, \apj, 376, 540

\bibitem[\protect\citeauthoryear{{Spangler} \& {Gwinn}}{{Spangler} \& {Gwinn}}{1990}]{spanglergwinn1990}
{Spangler} S.~R., {Gwinn} C.~R. 1990, \apjl, 353, L29

\bibitem[\protect\citeauthoryear{{Stinebring}, {McLaughlin}, {Cordes},
  {Becker}, {Goodman}, {Kramer}, {Sheckard} \& {Smith}}{{Stinebring}
  et~al.}{2001}]{stinebringetal2001}
{Stinebring} D.~R.,  {McLaughlin} M.~A.,  {Cordes} J.~M.,  {Becker} K.~M.,
  {Goodman} J.~E.~E.,  {Kramer} M.~A.,  {Sheckard} J.~L.,    {Smith} C.~T.,
  2001, \apjl, 549, L97

\bibitem[\protect\citeauthoryear{{Walker}, {de Bruyn} \& {Bignall}}{{Walker}
  et~al.}{2009}]{walkerdebruynbignall2009}
{Walker} M.~A.,  {de Bruyn} A.~G.,    {Bignall} H.~E.,  2009, \mnras, 397, 447

\bibitem[\protect\citeauthoryear{{Walker}, {Koopmans}, {Stinebring} \& {van
  Straten}}{{Walker} et~al.}{2008}]{walkeretal2008}
{Walker} M.~A.,  {Koopmans} L.~V.~E.,  {Stinebring} D.~R.,    {van Straten} W.,
   2008, \mnras, 388, 1214

\bibitem[\protect\citeauthoryear{{Walker}, {Melrose}, {Stinebring} \&
  {Zhang}}{{Walker} et~al.}{2004}]{walkeretal2004}
{Walker} M.~A.,  {Melrose} D.~B.,  {Stinebring} D.~R.,    {Zhang} C.~M.,  2004,
  \mnras, 354, 43

\end{thebibliography}
\end{document}